\documentclass[dvips]{arxstspdf}
\usepackage{flushend}
\usepackage{stfloats}
\usepackage{graphics}

\volume{24}
\issue{3}
\pubyear{2009}
\firstpage{319}
\lastpage{327}
\doi{10.1214/09-STS303}

\begin{document}
\begin{frontmatter}

\title{Inference and Modeling with Log-concave~Distributions}
\runtitle{Log-concave distributions}

\begin{aug}
\author[a]{\fnms{Guenther} \snm{Walther}\ead[label=e1]{gwalther@stanford.edu}\corref{}}
\runauthor{G. Walther}

\affiliation{Stanford University}

\address[a]{Guenther Walther is Associate Professor,
Department of Statistics, Stanford University, 390 Serra Mall,
Stanford, California 94305, USA
\printead{e1}.}

\end{aug}

%
\begin{abstract}
Log-concave distributions are an attractive choice for modeling
and inference, for several reasons: The~class of log-concave distributions
contains most of the commonly used parametric distributions
and thus is a rich and flexible nonparametric class of distributions.
Further, the MLE exists and can be computed with readily available
algorithms. Thus, no tuning parameter, such as a bandwidth, is necessary
for estimation. Due to these attractive properties, there has been
considerable recent research activity concerning the theory and applications
of log-concave distributions. This article gives a review of these
results.
\end{abstract}

%

\end{frontmatter}
%

\section{Introduction} \label{introduction}

There has been considerable recent activity in the area of inference
under shape constraints, that is, inference about a (say) function $f$
under the constraint that $f$ satisfies certain qualitative
properties, such as monotonicity or convexity on certain subsets
of its domain. This approach is appealing for two main reasons:
First, such shape constraints are sometimes direct consequences of the
problem under investigation (see, e.g., Hampel, \citeyear{Ha1987}, or Wang et al., \citeyear{WaWoWaMaOl2005}),
or they are at least plausible in many problems.
It is then desirable that~the result of the inference reflect this
fact. There is also the hope that imposing these constraints will
improve~the quality of the resulting estimator in some sense.
The~second reason is that alternative nonparametric estimators
such as, for example, kernel estimators, typically require the choice of
a tuning parameter such as a bandwidth. A~good choice for such a tuning
parameter is usually far from trivial and injects a certain amount of
subjectivity into the estimator.
In contrast, inference under shape constraints often results in an
explicit solution that does not depend on a tuning parameter.

In the context of density estimation, Grenander (\citeyear{Gr1956}) derived
the nonparametric maximum likelihood estimator of a density function
that is nonincreasing on a half-line. This estimator is given
explicitly by the left derivative of the least concave majorant
of the empirical distribution function. However, this result
does not carry over to the problem of estimating a unimodal density
with unknown mode, as then the nonparametric MLE does not exist;
see, for example, Birg\'{e} (\citeyear{Br1997}). Even if the mode is known, the estimator
suffers from inconsistency near the mode, the so-called spiking
problem; see, for example, Woodroofe and Sun (\citeyear{WoSu1993}). These results are
unfortunate since the constraint of unimodality is
cited as a reasonable assumption in many problems.

\begin{figure*}

\includegraphics{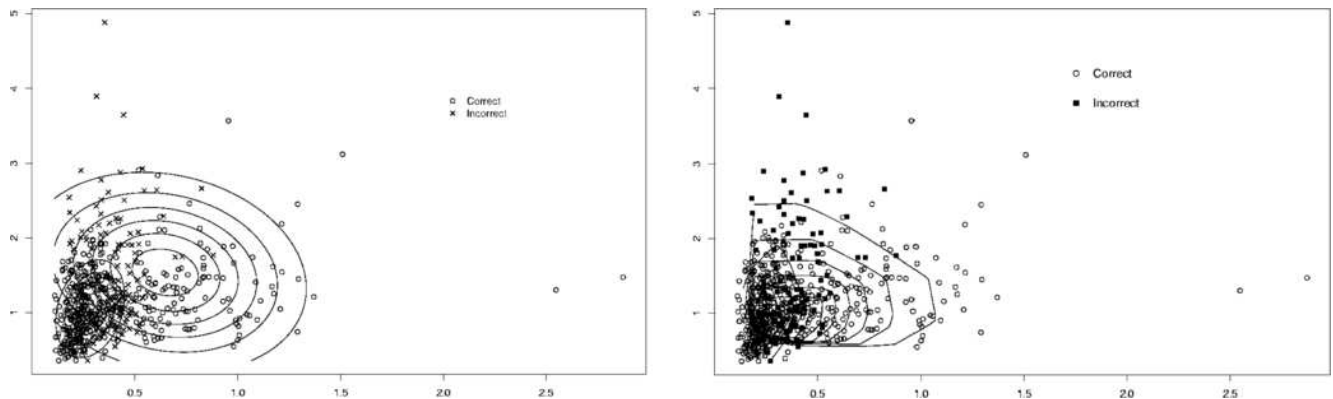}

\caption{Contour plot and misclassified instances from the Gaussian
EM algorithm (left) and the log-concave EM algorithm (right).
The~plots are from Cule, Samworth and Stewart (\protect\citeyear{CuSaSt2008}).}
\label{fig0}
\end{figure*}

It was argued in Walther (\citeyear{Wa2002}) that log-concave densities
are an attractive and natural alternative choice to the
class of unimodal densities:
The~class of log-concave densities is a subset
of the class of the unimodal densities, but it contains most of the
commonly used parametric distributions and is thus a rich and useful
nonparametric model. Moreover, it was shown in Walther (\citeyear{Wa2002})
that the nonparametric MLE of a univariate log-concave density exists
and can
be computed with readily available algorithms.

Due to these attractive properties, there has been considerable recent
research activity about the statistical properties of the MLE,
computational aspects, applications in modeling and inference,
as well as about the multivariate case.
As an example, Figure~\ref{fig0} shows a scatterplot of measurements
on 569 individuals from the Wisconsin breast cancer data set; see
Section~\ref{applications} for a more detailed description. The~data
were clustered using a two-component normal mixture model fitted with
the EM-algorithm; see, for example, Fraley and Raftery (\citeyear{FrRa2002}). The~contour
lines of the fitted normal components are shown in the left plot, while the
right plot shows the contour lines that obtain when the normal
MLE is replaced by the log-concave MLE in the EM algorithm. The~log-concave
MLE automatically adapts to the multivariate skewness of the data and
results in a superior clustering: Each observation is either a benign
or a malignant instance. These labels were not used for the fitting
but can be employed to assess the quality of the clustering.
The~EM algorithm with the log-concave MLE resulted in 121 misclassified
instances versus 144 for the Gaussian MLE.

This article gives an overview of recent results about inference
and modeling with the log-concave MLE.
Section~\ref{basics} gives some basic properties and applications
of log-concave distributions. Section~\ref{statprop} addresses the
MLE and its statistical properties. Computational aspects are
surveyed in Section~\ref{computation}, while Section~\ref{multivariate}
describes recent advances in the multivariate setting.
Section~\ref{applications} reviews applications
of the log-concave MLE for various modeling and inference
problems. Section~\ref{outlook} lists some open problems for
future work.

\section{Basic Properties and Applications of Log-concave Functions}
\label{basics}

A~function $f$ on $\mathbf{R}^d$ is log-concave if it is of the form
%
\begin{equation}\label{logconcave}
f(x) = \exp\phi(x),
\end{equation}
for some concave function $\phi  \dvtx   \mathbf{R}^d \rightarrow
[-\infty
,\infty)$.
A~pri\-me example is the normal density, where $\phi(x)$ is a quadratic
in $x$. Further, most common univariate parametric densities are log-concave,
such as the normal family, all gamma densities with shape parameter
$\geq1$,
all Weibull densities with exponent $\geq1$, all beta densities with both
parameters $\geq1$, the generalized Pareto and the logistic density;
see, for example, Marshall and Olkin (\citeyear{MaOl1979}).

\begin{figure*}[b]

\includegraphics{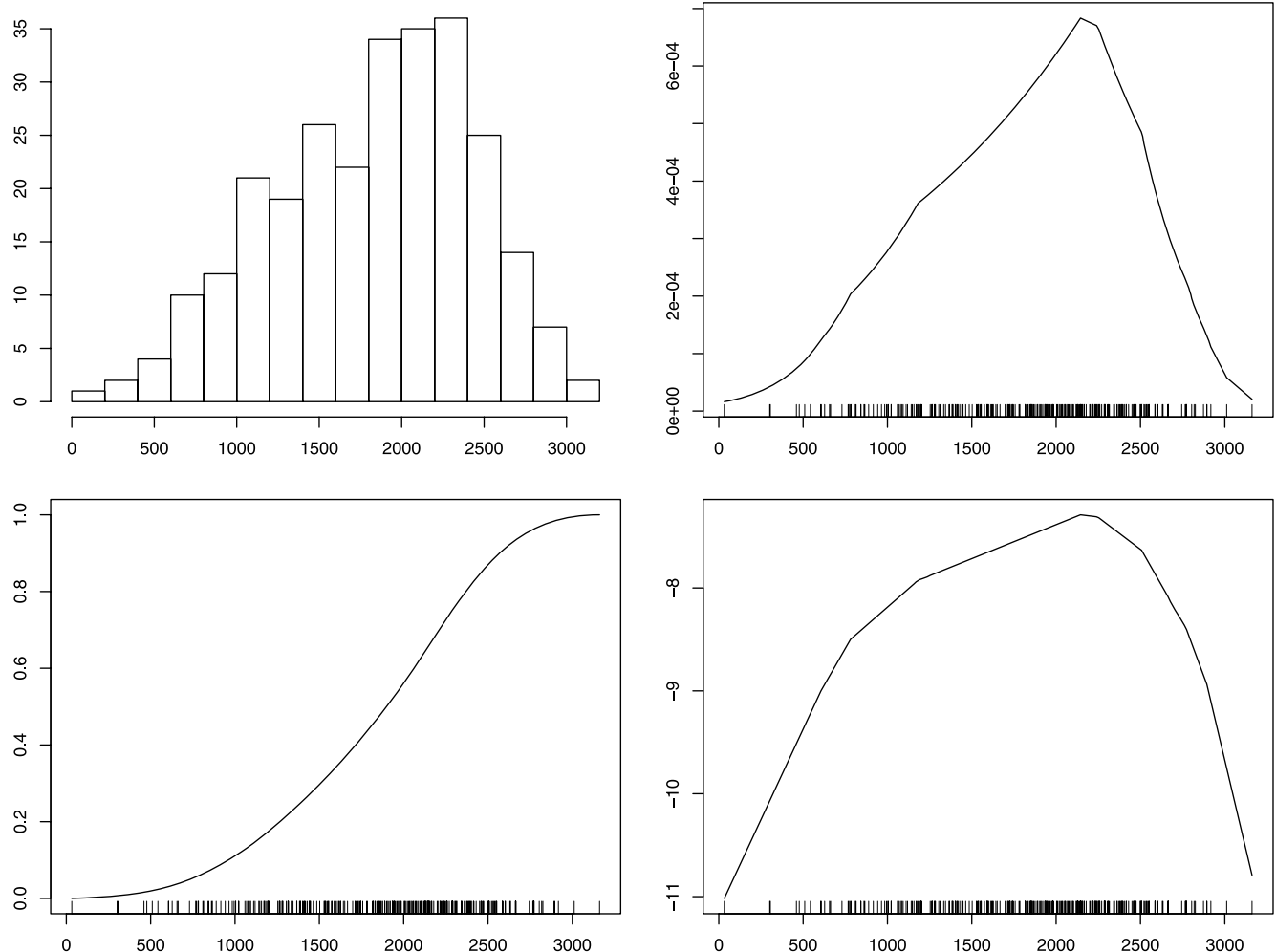}

\caption{The~histogram of $n=270$ flow cytometry data (top left), the
log-concave MLE $\hat{f}_n$ (top right), the estimated c.d.f. (bottom left),
and $\hat{\phi}_n=\log\hat{f}_n$ (bottom right).}
\label{fig1}
\end{figure*}

Log-concave functions have a number of properties that are desirable
for modeling: Marginal distributions, convolutions and product measures
of log-concave distributions are
again log-concave; see, for example, Dharmadhikari and Joag-Dev (\citeyear{DhJo1988}). Notably,
the first two properties are not true for the class of unimodal
densities.\footnote{Counterexamples are available from the author upon
request.} Log-concave distributions may be skewed, and this flexibility
is relevant in a number of applications; see, for example, Section~\ref
{applications}.
On the other hand, log-concave distributions necessarily have subexponential
tails and nondecreasing hazard rates; see, for example, Karlin (\citeyear{Ka1968})
and Barlow and Proschan (\citeyear{BaPr1975}).

There are several alternative characterizations and designations for
the class of univariate log-concave distributions: Ibragimov (\citeyear{Ib1956})
proved that these are precisely the distributions whose convolution
with a unimodal distribution is always unimodal; thus,
log-concave distributions are sometimes referred to as strongly
unimodal. Log-concave densities are also precisely the Polya frequency
functions of order 2, as well as precisely those densities $f$ for which
the location family $f_{\theta}(x):=f(x-\theta)$ has monotone likelihood
ratio in $x$; see Karlin (\citeyear{Ka1968}).

Log-concave distribution models have been found useful in economics
(see, e.g., An, \citeyear{An1995}, \citeyear{An1998}; Bagnoli and Bergstrom, \citeyear{BaBe2005} and Caplin and
Nalebuff, \citeyear{CaNa1991}), in
reliability theory (see, e.g., Barlow and Proschan, \citeyear{BaPr1975}) and in
sampling and nonparametric Bayesian analysis (see, e.g., Gilks and Wild,
\citeyear{GiWi1992};
Dellaportas and Smith, \citeyear{DeSm1993} and Brooks, \citeyear{Br1998}). Recent advances in
inference have led to fruitful applications of log-concave distributions
in other areas such as clustering, some of which
will be discussed in Section~\ref{applications}.

\section{Properties of the Nonparametric MLE}
\label{statprop}

If $X_1,\ldots,X_n$ are i.i.d. observations from a univariate
log-concave density (\ref{logconcave}), then the nonparametric MLE
exists, is unique, and is of the form $\hat{f}_n=\exp\hat{\phi}_n$,
where $\hat{\phi}_n$ is continuous and piecewise linear on $[X_{(1)},X_{(n)}]$
with the set of knots contained in $\{X_1,\ldots,X_n\}$, and
$\hat{\phi}_n=-\infty$ on $\mathbf{R}\setminus[X_{(1)},X_{(n)}]$;
see Walther (\citeyear{Wa2002}), Rufibach (\citeyear{Ru2006}) or Pal, Woodroofe and Meyer (\citeyear{PaWoMe2007}).
An example is plotted in Figure~\ref{fig1}.

Consistency of $\hat{f}_n$ with respect to the Hellinger metric was
established in Pal, Woodroofe and Meyer (\citeyear{PaWoMe2007}), while D\"{u}mbgen and Rufibach (\citeyear{DuRu2009})
provide results on the uniform consistency on compact subsets of the
interior of the support: If $\phi$ belongs to a H\"{o}lder class with
exponent $\beta\in[1,2]$, then $\hat{\phi}_n$ and\break $\hat{f}_n$
are uniformly consistent with rate\break $O_p ((\log n/n)^{\beta
/(2\beta+1)}
)$. Thus, in the typical case $\beta=2$, $\hat{f}_n$ converges
uniformly with rate $O_p ((\log n/n)^{2/5})$.
It is known that these rates are optimal even if $\beta$ were
known. This establishes that the nonparametric MLE adapts
to the unknown local smoothness of $f$, at least for $\beta\in[1,2]$.
Further, under some regularity conditions, the c.d.f. $\hat{F}_n$ of
$\hat{f}_n$
is asymptotically equivalent to the empirical c.d.f. $\mathbb{F}_n$: If
$\beta
>1,$ then
$|\mathbb{F}_n-\hat{F}_n|$ is of order $o_p(n^{-1/2})$ uniformly over compact
subsets of the interior of the support. Moreover,
$\mathbb{F}_n- n^{-1} \leq\hat{F}_n \leq\mathbb{F}_n$ on the set
of knots of $\hat
{\phi}_n$.
The~resulting uniform $\sqrt{n}$-consistency of $\hat{F}_n$
outperforms, for example, c.d.f.s of kernel estimators using a
nonnegative kernel with optimally
chosen bandwidth.
While empirical evidence suggests
that $\hat{f}_n$ performs well over the whole line, establishing the
corresponding theoretical results is still an open problem.

Balabdaoui, Rufibach and Wellner (\citeyear{BaRuWe2009})\break derive the pointwise limiting distributions
of\break $n^{k/(2k+1)}(\hat{f}_n(x_0)-f(x_0))$, $n^{(k-1)/(2k+1)}(\hat{f}_n'(x_0)
-\break f'(x_0))$, and likewise for $\hat{\phi}_n$ and $\hat{\phi}_n'$, where
$k$ is the smallest integer such that $\phi^{(k)}(x_0) \neq0$. They show
that these limiting distributions depend on the ``lower invelope'' of an
integrated Brownian motion process minus a drift term that depends
on $k$.

\section{Computational Aspects}
\label{computation}

Maximizing the log-likelihood function under the constraint
$\int\exp\phi(x)\, dx = 1$ is equivalent to maximizing
$\sum_{i=1}^n \phi(X_i) - n \int\exp\phi(x)\, dx$ over the
set of all concave functions $\phi$; see Silverman (\citeyear{Si1982}).
Due to the piecewise linear form of the solution $ \hat{\phi}$,
one can write this as a finite-dimensional optimization problem
as follows: For the ordered data $x_1<\cdots<x_n$ write
$\phi_1:=\phi(x_1)$ and denote the slope between $x_{i-1}$ and $x_i$
by $s_i:=(\phi(x_i)-\phi(x_{i-1}))/(x_i-x_{i-1})$, $i=2,\ldots,n$.
Then the optimization problem is to maximize
\begin{eqnarray*}
\hspace*{-4pt}&&\Psi_n (\phi_1,s_2,\ldots,s_n)\hspace*{5pt}\\
\hspace*{-4pt}&&\quad  = n\phi_1 +\sum_{i=2}^n(n-i+1)(x_i-x_{i-1})s_i \hspace*{4pt}\\
\hspace*{-4pt}& &\qquad {} -n\exp(\phi_1) \sum_{i=2}^n
\Biggl(\exp\Biggl( \sum_{k=2}^i(x_k-x_{k-1})s_k\Biggr)\hspace*{4pt}\\
\hspace*{-4pt}&&\hspace*{97pt}{} - \exp\Biggl( \sum_{k=2}^{i-1} (x_k-x_{k-1})s_k\Biggr)\Biggr)\Big/
{s_i}\hspace*{4pt}
\end{eqnarray*}
under the constraint that the vector $(\phi_1,s_2,\ldots,s_n)$
belongs to the
cone $\mathcal{C}_n:=\{ {\bf y} \in\mathbf{R}^n\dvtx  y_2 \geq\cdots\geq
y_n\}$.
$\Psi_n$ is a concave function on $\mathbf{R}^n$ which needs to be
maximized over
the convex cone $\mathcal{C}_n$. This is precisely the type of problem for which
the Iterative Convex Minorant Algorithm (ICMA) was developed; see Groeneboom
and Wellner (\citeyear{GrWe1992}) and Jongbloed\break (\citeyear{Jo1998}). The~key idea of that algorithm is
to approximate the concave function locally around the current candidate
solution by a quadratic form, which is then maximized by a Newton
procedure over the cone by using the pool-adjacent-violators algorithm.
This procedure is then iterated to the final solution. Walther (\citeyear{Wa2002}),
Pal, Woodroofe and Meyer (\citeyear{PaWoMe2007}) and Rufibach (\citeyear{Ru2007}) successfully employ the ICMA for
this problem.
The~last reference gives a very detailed description of the algorithm and
also compares the ICMA to several other algorithms that can be used
for this problem, such as an interior point method; see, for example,
Terlaky and Vial (\citeyear{TeVi1998}). The~ICMA shows a clearly superior performance
in these simulation studies. Recently, D\"{u}mbgen, H\"{u}sler and Rufibach (\citeyear{DuHuRu2007}) have
computed the log-concave MLE with an active set algorithm; see,
for example, Fletcher (\citeyear{Fl1987}). Active set algorithms have the attractive
property that they find the solution in finitely many steps, while the
iterations of the ICMA have to be terminated by a stopping criterion.
It appears that the active set algorithm provides the most efficient method
for computing the MLE to date. Both the ICMA and the active set algorithm
for computing the log-concave MLE are available with the $\mathbf{R}$ package
``\texttt{logcondens},'' which is accessible from ``\texttt{CRAN}.'' An alternative
way to compute the MLE with convex programming algorithms is described in
Koenker and Mizera (\citeyear{KoMi2008}).

Another advantage of the log-concave MLE $\hat{f}_n$
is that sampling from $\hat{f}_n$
is quite straightforward: First, compute the c.d.f. $\hat{F}_n$ at the
ordered sample $x_1,\ldots,x_n$ by integrating the piecewise exponential
function $\hat{f}_n$. Next, generate a random index $J \in\{2,\ldots
,n\}$
with $P(J=j)=\hat{F}_n(x_j)-\hat{F}_n(x_{j-1})$. Then generate
$U \sim U[0,1]$ and set $\Theta:=\hat{\phi}_n(x_J)-\hat{\phi}_n(x_{J-1})$.
If $\Theta\neq0,$ set $V:=\log(1+(\exp(\Theta)-1)U)/\Theta$,
otherwise set $V:=U$. Then $X:=x_{J-1}+(x_J-x_{J-1})V$ has density~$\hat{f}_n$.

\section{The~Multivariate Case}
\label{multivariate}

\begin{figure*}

\includegraphics{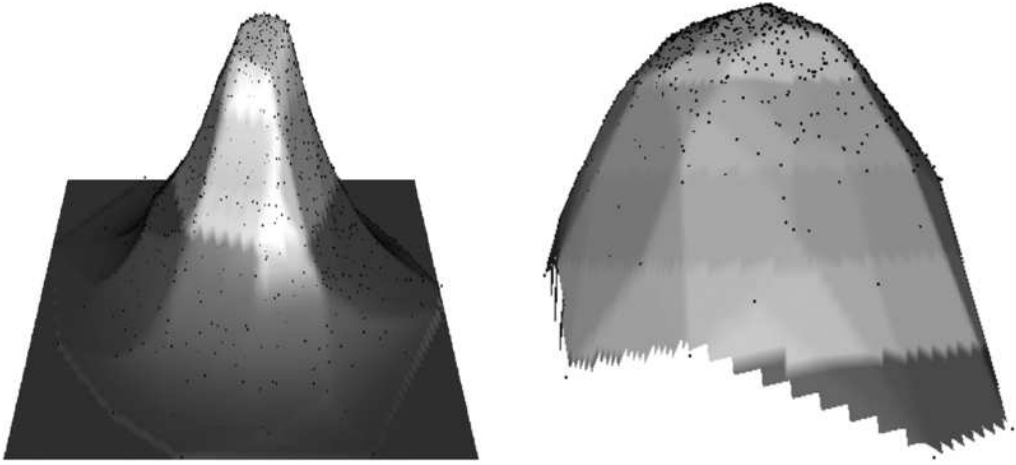}

\caption{The~MLE $\hat{f}_n$ (left) and $\hat{\phi}_n=\log\hat
{f}_n$ (right)
for $n=1000$ observations (plotted as dots) from a standard bivariate
normal distribution. The~plots are from Cule, Samworth and Stewart (\protect\citeyear{CuSaSt2008}).}
\label{fig2}
\end{figure*}

The~definition of a log-concave density does not depend
on the underlying dimension; see (\ref{logconcave}). The~fact that the
MLE does not require the choice of a tuning parameter makes its
use even more attractive in a multivariate setting, where, for example,
a kernel estimator requires the difficult choice of a bandwidth matrix.
The~structure of the multivariate MLE is analogous to the univariate
case; see, for example, Cule, Samworth and Stewart (\citeyear{CuSaSt2008}): The~support of the MLE is
the convex
hull of the data, and there is a triangulation of this convex hull such
that $\log\hat{f}_n$ is linear on each simplex of the triangulation.
Figure~\ref{fig2} depicts an example for two-dimensional data.
The~multivariate MLE has already shown promise in a number of
applications; see Section~\ref{applications}.

The~computation of the MLE requires an approach that is different from
the univariate setting, as the multivariate piecewise linear
structure of $\log\hat{f}_n$ does not allow to write this
optimization problem in terms of a simple ordering of the slopes.
Cule, Samworth and Stewart (\citeyear{CuSaSt2008}) show how the MLE can be computed by solving a
nondifferentiable convex optimization problem using Shor's $r$-algorithm;
see Kappel and Kuntsevich (\citeyear{KaKu2000}). Cule, Samworth and Stewart (\citeyear{CuSaSt2008}) report a robust
and accurate performance of this algorithm, which they implemented
in the \texttt{R} package \texttt{LogConcDEAD}; see Cule, Gramacy and Samworth (\citeyear{CuGrSa2009}).
However, the computation time increases
quickly with sample size and dimension. Cule, Samworth and Stewart (\citeyear{CuSaSt2008}) report
computation times of about 1 sec for $n=100$ observations in two
dimensions, to 37 min for a sample of size $n=1000$ in four
dimensions. It is therefore desirable to develop faster algorithms
for this problem.

Cule, Samworth and Stewart (\citeyear{CuSaSt2008}) investigate the finite sample performance of
the multivariate MLE via a simulation study. They compare the mean
integrated squared error of the MLE with that of a kernel estimator
with Gaussian kernel and a bandwidth that is either chosen
to minimize the mean integrated squared error (using knowledge about
the density that would not be available in practice) or determined
by an empirical bandwidth selector based on least squares cross
validation. The~MLE outperforms both of these estimators except for
small sample sizes, and the improvement can be quite dramatic.
On the other hand, in view of the work of Birg\'{e} and Massart (\citeyear{BiMa1993}),
it seems unlikely that the MLE will achieve optimal rates of
convergence in dimensions $d>4$, due to the richness of the class
of concave functions.
It would thus be helpful to have theoretical results about the performance
of the multivariate MLE. Deriving such results is an open problem.

\section{Applications in Modeling and Inference}
\label{applications}

One of the most fruitful applications of log-concave distributions has
been in the area of clustering. A principled and successful approach
to assign the observations to clusters is via the mixture model
$f(x)=\sum_{m=1}^k \pi_m f_m(x)$, where the mixture proportions $\pi_m$
are nonnegative and sum to unity, and the component distributions $f_m$
model the conditional density of the data in the $m$th cluster; see,
for example,
McLachlan and Peel (\citeyear{McPe2000}). Typically one assumes a parametric formulation
$f_m(x)=f(\theta_m,x)$ for the component distributions, such as the normal
model; see, for example, Fraley and Raftery (\citeyear{FrRa2002}). Then the EM
algorithm provides
an elegant solution to fit the above mixture model and to assign the data
to one of the $k$ components: The~EM algorithm iteratively assigns
the data based on the current maximum likelihood estimates of the component
distributions, and then updates those estimates $\hat{\pi}_m, \hat
{\theta}_m$
based on these assignments. An important advantage of using a mixture
model for clustering is that it provides not only an assignment of the
data to the $k$ components, but also a measure of uncertainty for this
assignment via the posterior probabilities that the $i$th observation
belongs to the $m$th component: $\hat{\pi}_m \hat{f}_m(X_i)/
\sum_{j=1}^k \hat{\pi}_j \hat{f}_j(X_i)$.

\begin{figure*}[b]

\includegraphics{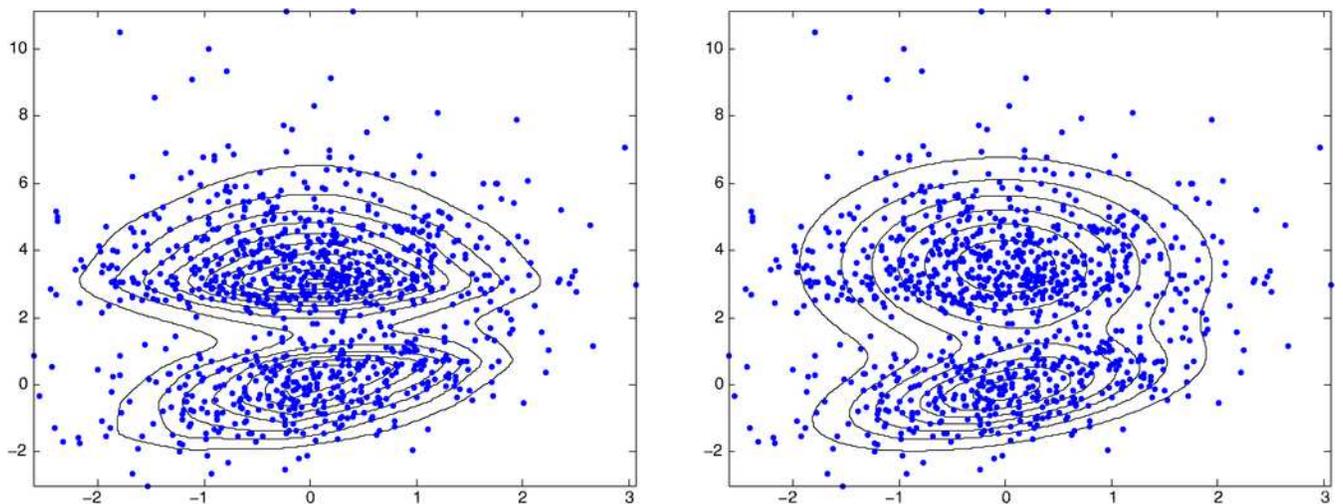}

\caption{Contours of the estimated model obtained from the log-concave EM
algorithm of Chang and Walther (\protect\citeyear{ChWa2007}) (left) and from the Gaussian
EM algorithm (right) based on the plotted observations. The~underlying
distribution has a skewed (shifted gamma) distribution in the $y$-direction
of the top component. The~plots are from Chang and Walther (\protect\citeyear{ChWa2007}).}
\label{fig3}
\end{figure*}

A~disadvantage of this approach is that it depends on the parametric
formulation in several important ways: If the parametric model is misspecified,
then the accuracy of the clustering may deteriorate and the measure of
uncertainty may be considerably off. For some data, such as those
in Figure~\ref{fig1}, no appropriate parametric model may be available.
Another disadvantage is that each parametric model requires a different
implementation of the EM algorithm based on certain theoretical
derivations; see, for example, McLachlan and Krishnan (\citeyear{McKr1997}).

Therefore, it is desirable to have an EM-type clustering algorithm
with nonparametric component distributions. This would allow for
a universal software implementation with flexible component distributions.
As was expounded in Sections~\ref{introduction} and~\ref{basics},
the class of log-concave distributions provides a flexible model,
and, moreover, the MLE exists. Thus, one may attempt to mimic the
EM-type clustering algorithm that works so well in the parametric
context. This idea was successfully carried out in Chang and Walther (\citeyear{ChWa2007})
and in Cule, Samworth and Stewart (\citeyear{CuSaSt2008}). In related work,
Eilers and Borgdorff (\citeyear{EiBo2007}) use a nonparametric
smoother in place of the log-concave MLE in the M-step, with a penalty term
that moves the estimate toward a log-concave function.
Chang and Walther (\citeyear{ChWa2007}) report a clear improvement compared
to the parametric EM algorithm when the parametric model is not
correct, and a performance that is almost similar to the Gaussian
EM algorithm in the case where the
Gaussian model is correct. Thus, the use of log-concave component
distributions provides a flexible methodology for clustering, and this
flexibility does not entail any noticeable penalty in the special
case where a parametric model is appropriate.

Chang and Walther (\citeyear{ChWa2007})
also consider a multivariate extension by modeling each component
distribution with log-concave marginals and a normal copula for
the dependence structure. This
simple multivariate extension avoids the more challenging task of
estimating a multivariate log-concave density, but it is flexible enough
for many situations. Figure~\ref{fig3} compares the fitted components
with those for the Gaussian model for simulated bivariate data.
The~log-concave model automatically
picks up the skewness in the $y$-direction and results in a noticeably
improved error rate for the clustering; see Chang and Walther (\citeyear{ChWa2007})
for details.\looseness=1

\begin{figure*}
\begin{tabular}{c}

\includegraphics{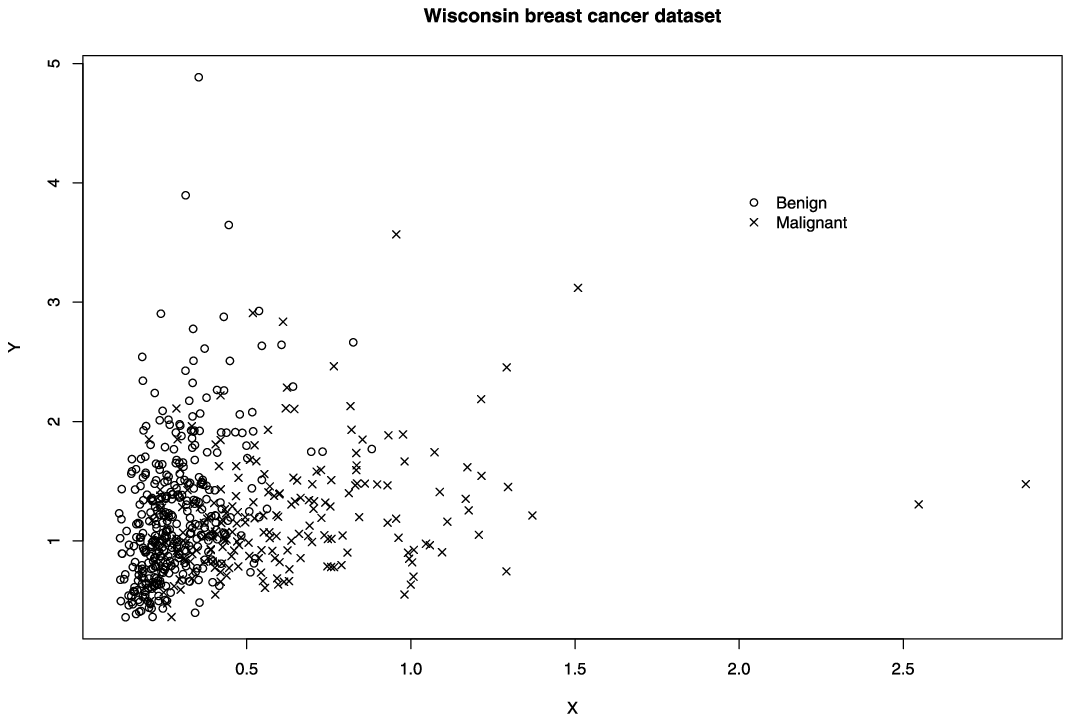}
\\[6pt]

\includegraphics{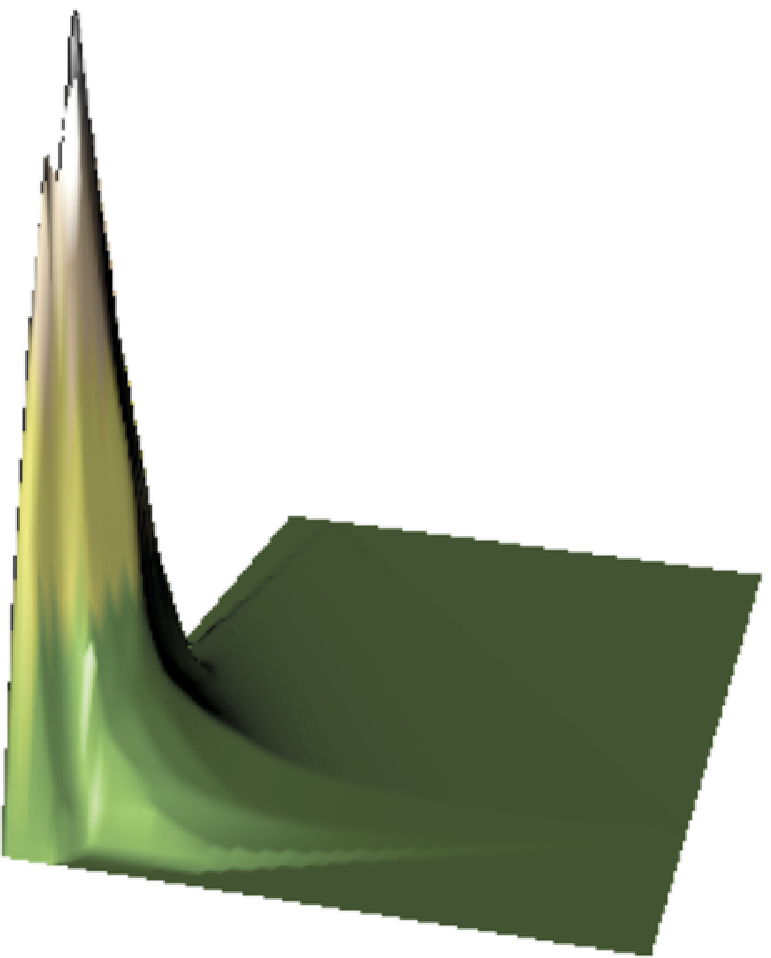}

\end{tabular}
\caption{The~Wisconsin breast cancer data (top), with benign cases
as open circles and malignant cases as crosses. The~bottom plot
shows the fitted mixture distribution from the log-concave EM algorithm.
The~plots are from Cule, Samworth and Stewart (\protect\citeyear{CuSaSt2008}).}
\label{fig4}
\end{figure*}

Cule, Samworth and Stewart (\citeyear{CuSaSt2008}) extend this approach by using the
multivariate log-concave MLE for each component. They apply
the log-concave EM algorithm to the Wisconsin breast cancer data
of Street et~al. (\citeyear{Stetal1993}) and obtain only 121 misclassified instances
compared to 144 with the Gaussian EM algorithm. Figure~\ref{fig4}
shows a scatterplot of the data and the fitted log-concave mixture.
The~contour plots of the fitted components from the Gaussian EM algorithm
and the log-concave EM algorithm are given in Figure~\ref{fig0}.

Developing principled methodology for selecting an appropriate number
of components is an open problem. Methodology for testing for the
presence of mixing in the log-concave model is given by Walther (\citeyear{Wa2001})
and Walther (\citeyear{Wa2002}), where the latter approach uses the fact that
a log-concave mixture allows the representation $\exp(\phi(x) +
c \Vert x \Vert^2)$ for some $c \geq0$ and a concave function
$\phi$.

While log-concave distributions allow for flexible modeling,
the structure provided by a log-concave estimator has turned out
to result in advantageous properties in a number of
other inference problems:

D\"{u}mbgen and Rufibach (\citeyear{DuRu2009}) use the fact that the hazard
rate of a log-concave density is automatically monotone and
construct a simple plug-in estimator of the hazard rate which
is nondecreasing. Rates of convergence for $\hat{f}_n$ automatically
translate to rates for the hazard rate estimator.

M\"{u}ller and Rufibach (\citeyear{MuRu2009}) report an
improved performance for certain problems in extreme value theory
when employing a log-concave estimator.

D\"{u}mbgen, H\"{u}sler and Rufibach (\citeyear{DuHuRu2007}) show how the assumption of log-concavity
allows the estimation of a distribution
based on arbitrarily censored data using the EM algorithm. They
replace the log-likelihood function by a function that is linear
in $\phi$. This function can be interpreted as the conditional
expectation of the log-likelihood function given the available data
and represents the E-step in the EM algorithm. The~M-step consists
of maximizing this function using the active set algorithm described
in Section~\ref{computation}.

Balabdaoui, Rufibach and Wellner (\citeyear{BaRuWe2009}) investigate the mode of $\hat{f}_n$ as an
estimator of the mode of $f$. Estimation of the mode of a unimodal density
has received considerable attention in the literature. Typically,
some choice of bandwidth or tuning parameter is required due to
the problems with the MLE of a univariate density described in
Section~\ref{introduction}. The~MLE of a log-concave density does
not suffer from this problem and provides an estimate of the
mode as a by-product. Balabdaoui, Rufibach and Wellner (\citeyear{BaRuWe2009}) establish the limiting
distribution of this estimator and show that the estimator is
optimal in the asymptotic minimax sense.

\section{Summary and Future Work}
\label{outlook}

Log-concave distributions constitute a flexible nonparametric class
which allows modeling and inference without a tuning parameter.
The~MLE has favorable theoretical performance properties and
can be computed with available algorithms. These advantageous
properties have resulted in tangible improvements in a number of
relevant problems, such as in clustering and when handling censored
data.

As for future work, there is clearly the potential for similar
improvements in a host of other problems, such as regression (see, e.g.,
Eilers, \citeyear{Ei2005}) or Cox regression under shape constraints on the hazard
rate.
Further, it would be useful to study the consequences of model
misspecification.
For example, the mode of the log-concave MLE is a useful tool for data analysis.
It would thus be interesting to investigate how far off this mode
can be from the population mode in the case where the population distribution
is unimodal but not log-concave.
The~outstanding performance of the multivariate
MLE reported in the simulation studies in Cule, Samworth and Stewart (\citeyear{CuSaSt2008})
lends importance to
a theoretical investigation of its convergence properties.
Finally, it would be desirable to develop faster algorithms for
computing the multivariate MLE.

For modeling with heavier, algebraic tails, it may be of interest to
consider the more general class of $\rho$-concave densities; see Avriel (\citeyear{Av1972}),
Borell (\citeyear{Bo1975}) and Dharmadhikari and Joag-Dev (\citeyear{DhJo1988}). First results about
nonparametric estimation and computational issues in this class were
obtained in Koenker and Mizera (\citeyear{KoMi2008}) and Seregin (\citeyear{Se2008}).

\section*{Acknowledgments}

 Thanks to Kaspar Rufibach and a referee
for comments and
several references, to Richard Samworth for providing figures,
and to Jon Wellner for bringing the
work of Arseni Seregin to my attention.
Work supported by NSF Grant DMS-05-05682 and NIH Grant
1R21AI069980.

\end{document}